\documentclass[%
 reprint,
nofootinbib,
 amsmath,amssymb,
 aps,
 prl,
]{revtex4-1}

\usepackage{float}
\usepackage[colorlinks = true, citecolor = blue, linkcolor = blue ]{hyperref}
\usepackage{graphicx}
\usepackage{dcolumn}
\usepackage{bm}

\newcommand{\jcap}{JCAP}

\newcommand{\apjl}{ApJL}
\newcommand{\apjs}{ApJS}

\newcommand{\mnras}{MNRAS}
\newcommand{\aap}{A\&A}

\newcommand{\pun}{M_{\odot} \rm km \ s^{-1}}
\newcommand{\pd}{p_{\rm dep}}
\newcommand{\pin}{P_{\rm in}}
\newcommand{\pcr}{P_{\rm CR}}
\newcommand{\esn}{E_{\rm SN}}
\newcommand{\rsh}{R_{\rm sh}}
\newcommand{\vsh}{V_{\rm sh}}
\newcommand{\vf}{V_{\rm f}}
\newcommand{\vd}{V_{\rm}}
\newcommand{\nism}{ n_{\rm ISM}}
\newcommand{\pism}{ P_{\rm ISM}}
\newcommand{\rism}{ \rho_{\rm ISM}}
\newcommand{\Tism}{ T_{\rm ISM}}
\newcommand{\m}{ m_{\rm p}}
\newcommand{\xcr}{ \xi_{\rm CR}}

\begin{document}

\preprint{APS/123-QED}

\title{Effect of Cosmic Rays on the Evolution\\ and Momentum Deposition of Supernova Remnants}

\author{Rebecca Diesing}
 \author{Damiano Caprioli}%
\affiliation{%
 Department of Astronomy and Astrophysics, University of Chicago, Chicago, IL 60637, USA
}%

\date{\today}

\begin{abstract}
Using a semianalytical approach based on the thin-shell approximation, we calculate the long-term evolution of supernova remnants (SNRs) while also accounting for the cosmic rays (CRs) accelerated at their blast waves.
Our solution reproduces the results of state-of-the-art hydro simulations across the adiabatic and radiative stages for the gas-only case, and it predicts that typical CR acceleration efficiencies ($\approx 10\%$ ) can boost SNR momentum deposition by a factor of 2--3. 
This enhancement can become as large as an order of magnitude in environments in which the gas experiences more severe radiative losses.
This result may have a crucial impact on modeling the effect of supernova feedback on star formation and galaxy evolution. 

\end{abstract}

\maketitle

\section{\label{sec:level1}Introduction}
One of the most significant challenges in modeling galaxy formation and evolution is proper accounting for the effect on star formation history of feedback from active galactic nuclei (for the most massive galaxies), stellar winds, and supernova explosions. 
Feedback is crucial for quenching star formation and launching galactic winds, which drive baryons out and enrich the intergalactic medium in metals \cite[e.g.,][]{ht17}. 
Supernovae inject both energy and momentum into the interstellar medium (ISM) but they are not resolved in most galaxy simulations, in which a grid cell typically contains more mass than a supernova remnant (SNR) and the deposited thermal energy is quickly radiated away (\emph{overcooling}).
Therefore, SN feedback is effectively parametrized by subgrid models for momentum deposition \citep[e.g.,][]{agertz+13,hopkins+18, pfrommer+17,pillepich+18}, which---in order to produce galaxies consistent with observations---typically require an extra boost with respect to the momentum yield calculated with numerical simulations of single SNRs \citep{ih15, ko15, martizzi+15, wn15,gentry+17}. 
SN clustering may be able to increase momentum deposition with respect to a linear superposition of single events, but its role still needs to be quantitatively addressed \citep[e.g.,][]{gentry+18}.

In this Letter we study the evolution of SNRs, also including the contribution of the nonthermal particles (hereafter, cosmic rays, CRs) that are efficiently produced at SNR blast waves via diffusive shock acceleration \citep[e.g.,][]{bell78a, bo78}.
Kinetic simulations show that shocks can channel as much as 10--20\% of their bulk energy into CRs \citep{DSA,electrons}, consistent with SNR multiwavelength emission \cite[e.g.,][]{tycho,pionbump}.
We show that CRs can significantly affect the late-time SNR evolution and boost momentum deposition, thereby providing a physical motivation for the heuristic boost typically accounted for in galaxy simulations.

The  evolution of a typical SNR goes through several stages.
Initially, the SNR undergoes a phase of rapid expansion (\emph{ejecta-dominated stage}), in which the inertia of the swept-up ambient medium is much smaller than the mass of the SN ejecta, $M_{\rm ej}$.
When such inertia is no longer negligible, the SNR enters the \emph{Sedov stage} and expands adiabatically until the post-shock temperature drops below $\sim 10^6$ K, when the thermal gas cools rapidly due to forbidden atomic transitions. 
The SNR keeps expanding because its internal pressure still exceeds the ambient pressure ($P_{\rm in}> \pism$, \emph{pressure-driven snowplow}), but simulations show that  the deposited momentum quickly saturates to $\pd\sim 3\times 10^5 \pun$, with $M_\odot$ the solar mass \citep{cioffi+88, ih15, ko15, martizzi+15, wn15,gentry+18,gentry+17}.
Eventually, when $P_{\rm in}\approx \pism$, expansion is slow and driven by relic kinetic energy (\emph{momentum-driven snowplow}).

The effect of CRs on SNR evolution is twofold. 
(1) Acting as a relativistic fluid, CRs suffer less adiabatic loss than the thermal gas, so that at late times they dominate the internal pressure. 
(2) Most importantly, the CR energy is not radiated away during the snowplow phase, but rather continues to support SNR expansion.
The importance of CRs can be estimated with the following simple argument.
Denoting the fraction of the SNR bulk pressure converted to CR pressure at the shock with $\xcr$ and the initial SN kinetic energy with $\esn$, one has $\pcr\approx 3 \xcr \esn/(4\pi \rsh^3)$,  $\rsh$ being the SNR shock radius.
Momentum deposition may continue until $\pin \approx \pcr\approx \pism=k_B \nism\Tism$, where $\nism$ and $\Tism$ are the number density and temperature of the ambient medium, and $k_{\rm B}$ is the Boltzmann constant.
Introducing the ambient density $\rism \approx \mu\m\nism$, where $\m$ is the proton mass and $\mu\approx 1.4$ for 10\% helium abundance, the total momentum deposited is $\pd\approx M_{\rm f} \vf $, with $M_{\rm f} \approx \xcr \esn\rism/\pism$ and $\vf \approx \sqrt{\frac 53 k_{\rm B}\Tism/(\mu \m)}$ of the order of the ambient sound speed.
In numbers: 
\begin{equation}
\pd \approx 9.44\times 10^6 \frac{\xcr E_{\rm SN}}{10^{51} {\rm erg}}\left(\frac{\Tism}{8000\rm  K}\right)^{-\frac{1}{2}} M_{\odot} \rm km \ s^{-1}.
\label{eqn:pestimate}
\end{equation}

Physically speaking, in the presence of CRs, the radiative stage becomes similar to an adiabatic stage (with $\rsh\propto t^{2/5}$) with an effective SNR energy of $\sim \xcr\esn$.
Note that, if the ISM pressure were dominated by supersonic turbulence, $\vf$ would instead be comparable with the turbulent velocity, typically a few tens of km s$^{-1}$ and independent of $\Tism$.

This estimate hinges on the reasonable assumption that CRs remain confined in the SNR.  
$\gamma$-ray observations suggest that GeV CRs, which carry most of the pressure, are still present in middle-age or old SNRs \citep[e.g.,][]{gamma,efficiency}, even if higher-energy CRs may have already escaped \citep[e.g.,][]{escape, crspectrum}.
For a Bohm diffusion coefficient $D = cr_{\rm L}/3$ \citep{diffusion}, with $c$ the speed of light and $r_{\rm L}\approx 10^{12}$ cm the gyroradius of GeV CRs in a few $\mu$G magnetic field, the diffusion time on a distance comparable to the SNR shell thickness, $\Delta_{\rm sh}\sim$ a few pc, is in fact $\Delta_{\rm sh}^2/D\gtrsim  10^7$ yr,  longer than the SNR lifetime.

\section{\label{sec:level1}Method}
Both analytical (self-similar) and numerical solutions suggest that the evolution of a strong blast wave can be modeled in the so-called \emph{thin-shell approximation}, i.e., by assuming that most of the swept-up mass lies in a narrow layer, while the inner cavity of the SNR is filled with hot plasma \citep{Chernyi57, bs95,om88,bp04}. 
In this limit, the thin, dense shell stores the SNR kinetic energy, while the hot bubble contributes the internal pressure.
The standard thin-shell equations use different approximations in the adiabatic and radiative stages \citep[e.g.,][\S IIC]{bs95}; instead, we put forward a smooth, semianalytical solution that is valid from the beginning of the Sedov stage to the end of the pressure-driven snowplow stage, also including the pressure in CRs, the role of which has previously been accounted for during the Sedov stage only \citep[][\S  E2]{chevalier83,om88}.

\emph{Ejecta-dominated Stage---}In this stage, the post-shock shell is still forming, so we use the approximate analytical solution derived by \cite{tm99} for a SNR expanding in a uniform medium, where the SNR radius evolves as
\begin{equation}
\rsh(t) \approx \left[\left(\frac{0.25 M_{\rm ej}}{t^{2}E_{\rm SN}}\right)^\frac{3}{4} +\left(\frac{0.62\rism}{M_{\rm ej}}\right)^\frac{1}{2}\right]^{-\frac{2}{3}},
\end{equation}
and the shock velocity reads $\vsh(t)=d\rsh/dt$. 
This solution gives a smooth transition to the Sedov stage. It does not include the CR pressure, since CR acceleration is expected to 
ramp up during this time \citep[e.g.,][]{efficiency}.

\emph{Sedov stage---}When the the swept-up mass becomes comparable to $ M_{\rm ej} $,  one can use the thin-shell approximation  and write momentum conservation as \citep{bs95}:
\begin{equation}
\frac{d}{dt} \left(M\vd\right) = 4\pi\rsh^2( P_{\rm th}+P_{\rm CR}-\pism),
\label{eqn:pconservation}
\end{equation}
where $M(r)\equiv M_{\rm ej} + \frac{4\pi}{3}r^3\rism $ is the shell mass at radius $r$, $M\equiv M(\rsh)$, $\vd\equiv 2\vsh/(\gamma_{\rm eff} + 1)$ is the gas velocity immediately downstream of the shock, and $\gamma_{\rm eff}$ is the effective adiabatic index of the bubble.
For a mixture of thermal gas with $\gamma_{\rm th}=\frac{5}{3}$ and CRs with $\gamma_{\rm CR}=\frac{4}{3}$:
\begin{equation}
\gamma_{\rm eff} \equiv \frac{5+3w}{3(1+w)},\ \ w \equiv \frac{P_{\rm CR}}{P_{\rm CR} + P_{\rm th}},
\label{eq:gammaeff}
\end{equation}
where $w$ is the fraction of the total pressure in CRs \cite{chevalier83}.
Since $P_{\rm th}\simeq \frac{3}{4} \rism \vsh^2$ for a strong shock, $w\approx \frac{4}{3}\xcr$, where $\xcr$ is the CR acceleration efficiency defined above.

Introducing the SNR energy $E(\rsh)\equiv \frac{1}{2}M \vd^2+ E_{\rm th}+E_{\rm CR}$, and using $P_i/E_i=\gamma_i-1$, Eq.~\ref{eqn:pconservation} can be recast as:
\begin{equation}\label{eqn:integratedvelocity}
\begin{split}
& \vsh(\rsh) = \\ 
& =\frac{\gamma_{\rm eff}+1}{ M \rsh^{\lambda/2}}\left[
\int_{0}^{\rsh} dr r^\lambda M(r) \left(\frac{\lambda E(r) }{2r} - F_{\rm ISM}\right)\right]^{\frac{1}{2}},
\end{split}
\end{equation}
where $\lambda \equiv 6(\gamma_{\rm eff}-1)/(\gamma_{\rm eff}+1)$ and $ F_{\rm ISM} \equiv 4 \pi r^{2}\pism/(\gamma_{\rm eff}+1)$.

Finally, the SNR radius, $\rsh(t)$, can be calculated by using Eq.~\ref{eqn:integratedvelocity} and inverting 
\begin{equation}
t(\rsh) = \int_{0}^{\rsh}\frac{dr}{\vsh(r)} .
\end{equation}
During the Sedov stage $E(\rsh)=\esn$ and its partition into kinetic, thermal, and CR terms is fixed.

\emph{Pressure-driven snowplow stage---}When the post-shock temperature drops below $\sim 10^6$ K,  the shell becomes radiative and the SNR energy decreses as 
\begin{equation}\label{eq:radloss}
E(\rsh) = E_{\rm SN}  - \int_{R_{\rm rad}}^{\rsh} dr 2\pi\chi_{\rm th}\rism \vsh^2 \rsh^2,
\end{equation}
where $\chi_{\rm th}$ is the fraction of the energy flux across the shock, $\sim\frac{1}{2}\rism\vsh^3 $, that is immediately radiated away rather than added to the thermal gas.  
Typically, $\chi_{\rm th}\lesssim 0.7$, as calculated in the self-similar solution for an adiabatic stage with CRs \citep[table 6 of ref.][]{chevalier83}. 
While cooling begins during the Sedov stage and depends on the density/temperature profile \citep[e.g.,][]{draine11}, here we simply assume that radiative losses abruptly kick in at $\rsh=R_{\rm rad}$, chosen according to the simulations of \cite{ko15}.
The energy radiated away is subtracted from the thermal gas, so that $P_{\rm th}$ drops and the SNR is increasingly supported by $P_{\rm CR}$;
while $\gamma_{\rm eff}$ differs slightly between the hot bubble and the shell \citep{om88}, $\gamma_{\rm eff}\to \frac{4}{3}$ in both cases.  
Moreover, allowing $\gamma_{\rm eff}$ to vary between $\frac{4}{3}$ and $\frac{5}{3}$ introduces $\lesssim 10\%$ variations in $\pd$.
We stop the SNR evolution when the pressure inside the bubble equilibrates with that of the ISM, which returns a saturated $\pd = M  \vd$ (see Eq.~\ref{eqn:pconservation}).

The semianalytical solution proposed here is original in two respects: first, it accounts for the role of CRs in the radiative stage; 
second, the physically-motivated prescription for losses in Eq.~\ref{eq:radloss} smoothly follows the SNR evolution through the adiabatic and into the pressure-driven snowplow stage, unlike the standard thin-shell approximation \citep[see, e.g.,][]{ko15}. 
This formalism can be extended to more complex environments including pre-SN stellar winds \citep{gamma} and multiphase media \citep{bs95}.

\section{\label{sec:level1}Results}
\begin{figure}[ht]
\centering
\includegraphics[trim=5px 10px 0px 20px, clip, width=0.48\textwidth]{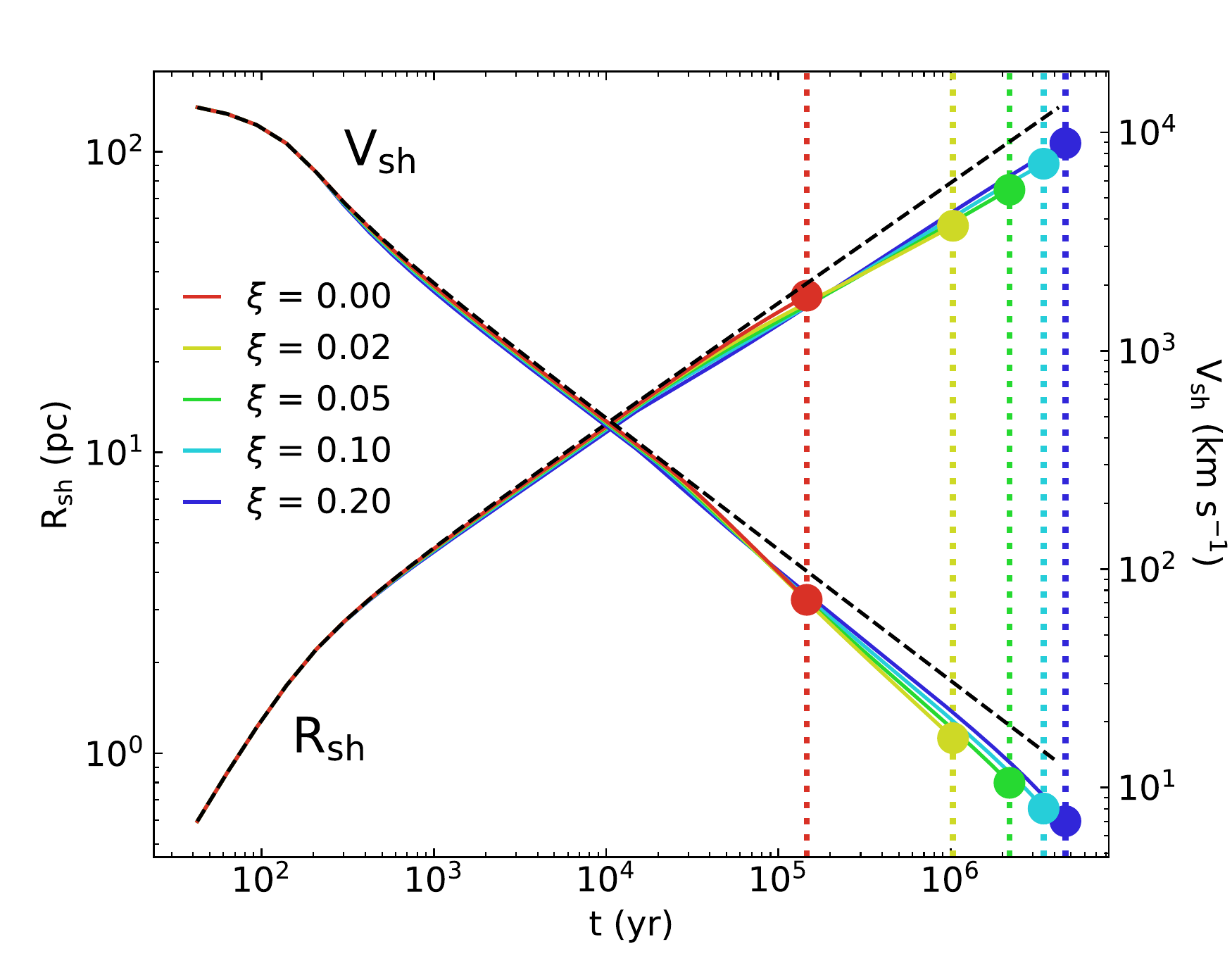}
\caption{Shock radius and velocity ($\rsh$ and $\vsh$)  as a function of time for various CR acceleration efficiencies, $\xcr$. 
In all cases, $E_{\rm SN} = 10^{51}$ erg, $M_{\rm ej} = 1$ $M_\odot$, $\nism =1 \rm cm^{-3}$, and $\Tism = 8000$ K. 
The dashed curve shows  the adiabatic solution with $\xcr$ = 0 \citep[][]{tm99}.
Circles indicate when the bubble and ambient pressures equilibrate and momentum deposition ceases.
The larger the CR acceleration efficiency, the longer the SNR evolution and the larger its final radius.}
\label{fig:radius_vs_time}
\end{figure}

SNR evolution with CR acceleration efficiencies $\xcr$ up to 20\% is plotted in Figure \ref{fig:radius_vs_time}. 
The presence of CRs slows the shock slightly during the Sedov stage due to the increase in the compressibility of the shell (lower $\gamma_{\rm eff}$), which reduces the size of the shell relative to the case with thermal gas alone \citep{chevalier83,om88}. 
However, the effect of CRs reverses in the radiative stage; since the energy in CRs cannot be radiated away, SNRs with CRs survive longer. 
For typical parameters, this latter effect dominates, meaning that CRs increase the amount of momentum deposited in the ISM.

\begin{figure}[ht]
\centering
\includegraphics[trim=5px 10px 20px 20px, clip, width= 0.48 \textwidth]{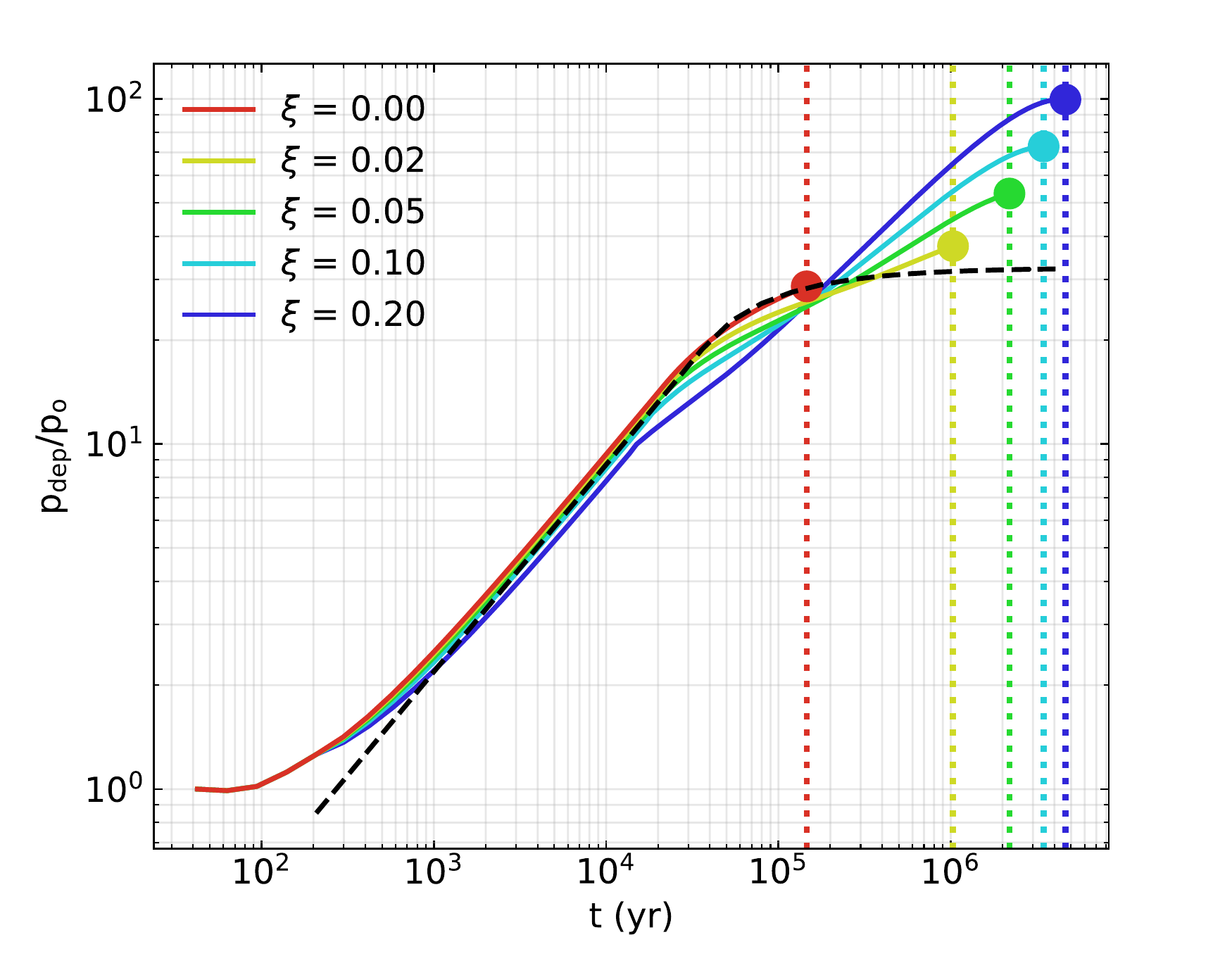}
\caption{Momentum deposited in the ISM, $\pd$, normalized to a reference initial momentum $p_0$, as a function of time for various CR acceleration efficiencies, $\xcr$, with the same parameters as in Figure \ref{fig:radius_vs_time}.
The gas-only case matches the fit to hydro simulations well \citep[][dashed lines]{ko15}; for typical values of $\xcr\gtrsim 0.1$, $\pd$ is boosted by a factor of 2 to 3. }
\label{fig:momentum_vs_time}
\end{figure}

Figure \ref{fig:momentum_vs_time} illustrates the net effect of CRs in regulating momentum deposition.
Whereas the $\xcr$ = 0 shock stalls at the onset of the pressure-driven snowplow, shocks with CRs continue to expand. 
As a result, CRs typically boost the momentum deposition by a factor of a few. 
For an acceleration efficiency of $\xcr = 10\%$, we find $\pd \approx 8\times 10^5 M_{\odot} \rm km s^{-1}$, in good agreement with the simple estimate in Eq.~\ref{eqn:pestimate}.
However, the energy deposited by the SNR in the presence of CRs is no more than 20\% larger than that of the gas-only case.  

\begin{figure}[h]
\centering
\includegraphics[trim=5px 16px 20px 20px, clip, width=0.48\textwidth]{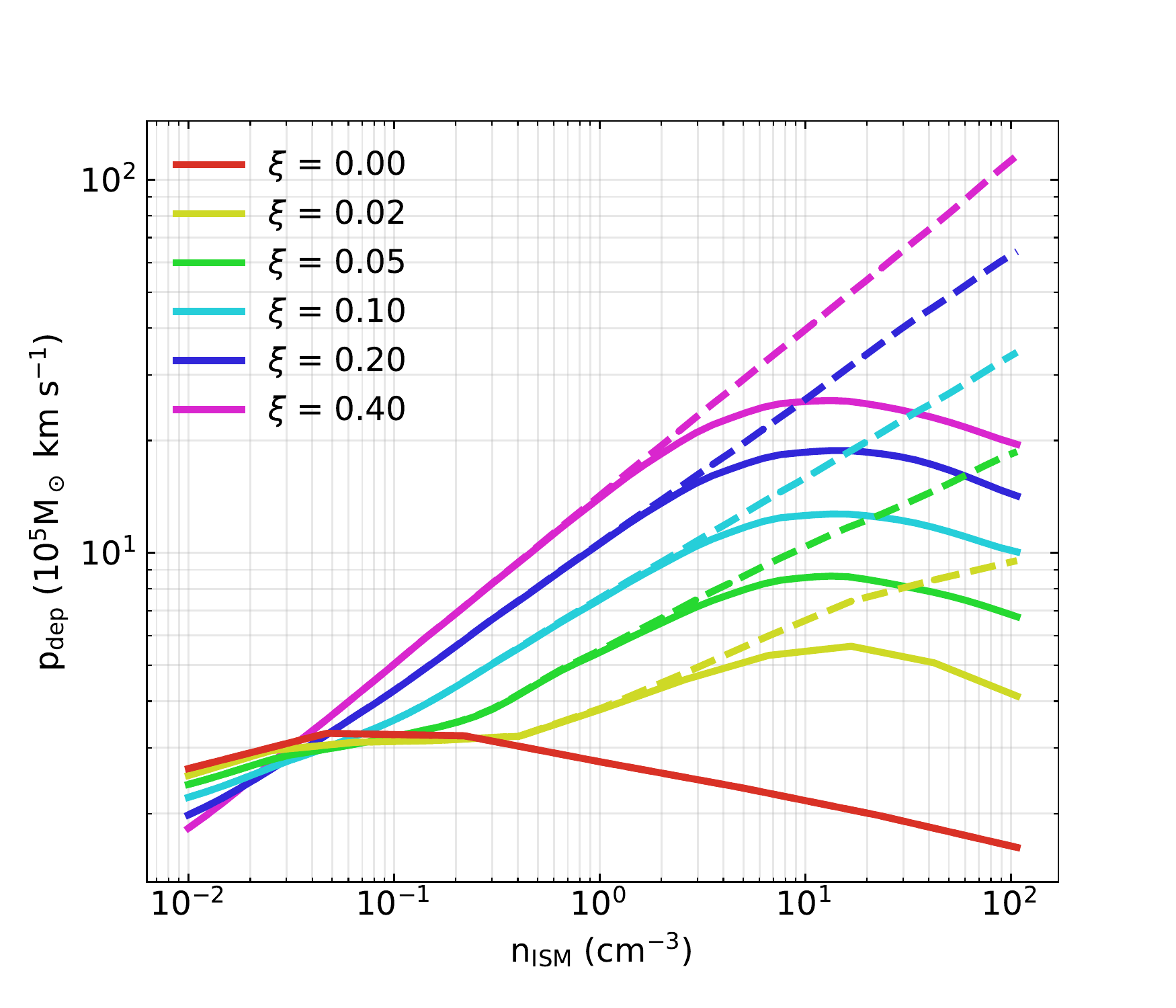}
\caption{Momentum deposited, $\pd$, for different values of $\xcr$, as a function of the ISM density when the ambient pressure $\pism\propto \nism \Tism$ is kept constant; all other parameters are fixed as in Figure \ref{fig:radius_vs_time}.
Solid and dashed lines correspond to cases with and without CR losses due to proton-proton and Coulomb collisions.
At large densities the boost due to the presence of CR saturates to a factor of 5--10 for $\xcr\gtrsim 0.1$.
\label{fig:momentum_vs_density_Pconst}}
\end{figure}

The effect of CRs becomes more pronounced in high ISM densities, where radiative losses are stronger.
Figure \ref{fig:momentum_vs_density_Pconst} shows $\pd$ for SNRs expanding in media with fixed pressure $\pism\propto \nism \Tism$ but different $\nism$.
An increase in density slows the evolution of the shell, meaning that it becomes radiative at an earlier time and, in the absence of CRs, is substantially shorter-lived.  For $\xcr = 0$ the result is a weak inverse relationship between ISM density and momentum deposited, $\pd\propto \nism^{-0.15}$, consistent with full hydro simulations \citep[e.g.,][]{martizzi+15, ko15}.
However, when CRs sustain shell expansion, increased density causes an increase in the swept-up mass that exceeds the corresponding decrease in the shell's velocity. 
The net effect is a positive relationship between ISM density and momentum deposited, $\pd\propto \sqrt{\nism}\propto 1/\sqrt{\Tism}$, as predicted by Eq.~\ref{eqn:pestimate} (dashed curves in Figure \ref{fig:momentum_vs_density_Pconst}). 

At large densities, CR energy losses due to inelastic proton--proton scattering and Coulomb interactions may not be negligible. 
We model such losses using the following rate that accounts for both effects \cite[e.g.,][]{pfrommer+17}:
\begin{equation} 
\Lambda_{\rm loss} \approx (7.44\mu+2.78)\times 10^{-16}n_{\rm ISM}\xcr\esn~\rm erg \ s^{-1};
\end{equation}
the corresponding curves for $\pd$ are shown as solid lines in Figure \ref{fig:momentum_vs_density_Pconst} and reported in Table \ref{tab:momentum_vs_density}.
CR losses are negligible for $\nism\lesssim 1$ cm$^{-3}$, but they limit $\pd$ for larger ISM densities; 
eventually, the boost due to CRs saturates to values $\gtrsim 5$ for $\nism\gtrsim10$ cm$^{-3}$.
Given that CRs are more important when radiative losses are more severe, we also expect a boost in momentum deposition when SNRs expand in a clumpy ISM. Where without CRs, $\pd$ is reduced due to rapid losses in the densest regions \citep[e.g.,][]{martizzi+15}, and when they expand out of rarefied circumstellar bubbles excavated by pre-SN stellar winds \citep[e.g.,][]{weaver+77,gamma}.

When SNRs expand into dense media, additional effects may become important.
First, the dynamics of shocks in partially-ionized media are nontrivially affected by the momentum and energy carried by neutral atoms, which are coupled to the thermal protons via charge-exchange  \citep{neutri1,neutri3}; 
the presence of neutrals, which requires a kinetic treatment and cannot be accounted for in hydro/MHD simulations, generally tends to make the shock weaker, reduce the post-shock temperature, and hence make radiative losses more important. 
In this case, including CRs should still increase momentum deposition with respect to the gas-only case. 
Second, ion-neutral damping \citep{kp69} may reduce the scattering of CRs and enhance their escape rate, even if a quantitative assessment of its relevance in realistic environments is still missing.
Third,  in turbulent molecular clouds, SNRs may dissolve when the shock speed becomes of order of the turbulent speed; this may reduce $\pd$ (see the discussion after Eq.~\ref{eqn:pestimate}). 

CR-loaded SNRs can sweep up more mass, leading to more prominent $\gamma$-ray emission due to the decay of neutral pions produced in nuclear interactions.
Middle-age or old (radiative) SNRs interacting with dense molecular clouds are often very bright in hadronic $\gamma$ rays  \citep[e.g.,][]{cs10, pionbump,acero+16}; 
the observed luminosities are typically explained with large efficiencies $\xcr\approx 0.1$ and prolonged CR confinement, which corroborates the assumptions of our calculation. 

The typical values $\xcr$ that pertain to strong shocks are in the range 0.1--0.2 \citep{DSA}, but reacceleration of preexisting CRs can lead to even larger efficiencies, for instance, in regions  rich in young stars with powerful winds and/or multiple SNe \citep{seeds}.
With this physical picture in mind, in Figure \ref{fig:momentum_vs_density_Pconst} we also consider the very efficient case $\xcr=0.4$.
For large densities $\nism\gtrsim 10$ cm$^{-3}$, the asymptotic boost in momentum deposition can be generally expressed as $\simeq 5 \sqrt{\xcr/0.1}$.

\begin{table}[h]
\resizebox{0.48\textwidth}{!}{\begin{tabular}{ c c c c c c c }
\hline \hline
$\nism \ (\rm cm^{-3})$ & \multicolumn{6}{c}{$\pd \ (10^5 M_{\odot} \rm km \ s^{-1})$} \\
\ & $\xi=0.00$ & $\xi=0.02$ & $\xi=0.05$ & $\xi=0.10$ & $\xi=0.20$ & $\xi=0.40$ \\
\hline
$10^{-2}$ & 2.64 & 2.53 & 2.39 & 2.21 & 1.98 & 1.83\\
$10^{-1}$ & 3.31 & 3.15 & 3.2 & 3.56 & 4.27 & 5.03\\
$10^0$ & 2.82 & 3.79 & 5.39 & 7.48 & 10.57 & 14.03\\
$10^1$ & 2.12 & 5.55 & 8.46 & 12.47 & 18.71 & 25.58\\
$10^2$ & 1.52 & 4.17 & 6.69 & 9.96 & 14.34 & 19.64\\
\hline \hline
\end{tabular}}
\caption{Momentum deposition for different $\xcr$ and $\nism$ with fixed ambient pressure $\pism\propto \nism \Tism$, as plotted in Figure \ref{fig:radius_vs_time}. CR losses are included.}
\label{tab:momentum_vs_density}

\end{table}
\section{\label{sec:level1}Conclusions}
We have presented the first semianalytical calculation ---in the thin-shell approximation limit--- of the evolution of a SNR throughout the adiabatic and radiative stages, while including the dynamical role of CRs accelerated at its forward shock. 
Despite the approximate treatment of the internal structure of the SNR, such an approach accurately reproduces the main features of hydro simulations without CRs. 
The presence of relativistic particles that, unlike the thermal gas, do not radiate efficiently, sustains a prolonged expansion of the shell (Figure \ref{fig:radius_vs_time}), which has the net effect of increasing the total momentum that each SN explosion can deposit in the ISM, $\pd$. 
For typical acceleration efficiencies of $\xcr\gtrsim 0.1$ \citep[e.g.,][]{DSA}, SNRs expanding into the warm ISM ($\Tism\simeq 10^4$ K and $\nism=1 \rm cm^{-3}$) deposit a {\bf factor of 2--3} more momentum than without CRs (Figure \ref{fig:momentum_vs_time}).
The impact of CRs on SNR evolution is stronger when radiation losses are more important, i.e., for dense and cold ambient gas. 
In such conditions, despite CRs losing energy due to inelastic proton--proton or Coulomb collisions, the boost in momentum deposition is even larger and reads $\simeq 5 \sqrt{\xcr/0.1}$. 

Simulations of galaxy formation that include SN feedback via subgrid models should account for the additional contribution of the CRs accelerated at SNR shocks; we quantify this contribution as a factor of a few to ten on top of the contribution of thermal gas alone, and provide the absolute values for $\pd$ in Table \ref{tab:momentum_vs_density}.
Note that this effect of CRs in shaping galaxies adds to that of the galactic-scale CR pressure gradient, which may play a role in launching galactic winds \citep[e.g.,][]{breitschwerdt+91,sb14, recchia+16, girichidis+16, farber+17, ruszkowski+17, recchia+18}.

\begin{acknowledgments}
The authors would like to thank A. Kravtsov, V. Semenov, E. Ostriker, R. Bandiera, T. Naab,  C. Pfrommer, C.A. Faucher-Gigu\'ere, C. Mc Kee, and E. Quataert for their valuable comments. 
This research was supported by NASA (grant NNX17AG30G to DC) and NSF (grant AST-1714658 to DC).
\end{acknowledgments}


\bibliography{DiesingCaprioli18}

\end{document}